# Resilience of coupled systems under deep uncertainty and dynamic complexity: An integrative literature review

Jannie Coenen, Vítor Vasconcelos, Heiman Wertheim, Marcel Olde Rikkert, Sophie Hadjisotiriou, Vittorio Nespeca, Tom Oreel, Rick Quax, Etiënne Rouwette, Vincent Marchau, and Hubert Korzilius*

Jannie Coenen, Radboud University, Institute for Management Research, Nijmegen, the Netherlands; jannie.coenen@ru.nl, orcid https://orcid.org/0000-0003-2577-3790

Vítor Vasconcelos, University of Amsterdam, Computational Science Institute, Amsterdam, the Netherlands; v.v.vasconcelos@uva.nl, orcid https://orcid.org/ 0000-0002-4621-5272

Heiman Wertheim, Radboud University Medical Centre, Department of Medical Microbiology, Nijmegen, the Netherlands, Heiman.wertheim@radboudumc.nl, orcid https://orcid.org/ 0000-0002-5003-5565

Marcel Olde Rikkert, Radboud University Medical Centre, Geriatrics Department, Nijmegen, the Netherlands; Marcel.OldeRikkert@radboudumc.nl, orcid https://orcid.org/ 0000-0003-1397-1677

Sophie Hadjisotiriou, Radboud University Medical Centre, Geriatrics Department, Nijmegen, the Netherlands; sophie.hadjisotiriou@radboudumc.nl, orcid https://orcid.org/0009-0001-1673-2340

Vittorio Nespeca, Leiden University, Faculty of Governance and Global Affairs, Leiden, the Netherlands, v.n.nespeca@fgga.leidenuniv.nl, orcid https://orcid.org/ 0000-0002-1691-0143

Tom Oreel, Erasmus University Rotterdam, Erasmus School of Social and Behavioural Sciences, Rotterdam, the Netherlands, oreel@essb.eur.nl, orcid https://orcid.org/0000-0001-5012-4878

Rick Quax, University of Amsterdam, Computational Science Institute, Amsterdam, the Netherlands, r.quax@uva.nl, orcid https://orcid.org/0000-0002-0299-0074

Etiënne Rouwette, Radboud University, Institute for Management Research, Nijmegen, the Netherlands, etienne.rouwette@ru.nl, orcid https://orcid.org/0000-0001-7178-8219

Vincent Marchau, Radboud University, Institute for Management Research, Nijmegen, the Netherlands, vincent.marchau@ru.nl, orcid https://orcid.org/0000-0002-8512-2386

Hubert Korzilius, Radboud University, Institute for Management Research, Nijmegen, the Netherlands, hubert.korzilius@ru.nl, orcid https://orcid.org/0000-0003-3728-6519, *corresponding author

# Resilience of coupled systems under deep uncertainty and dynamic complexity: An integrative literature review

**Abstract**
Resilience in coupled systems is increasingly critical in addressing global challenges such as climate change and pandemics. These systems show unpredictable behaviour due to dynamic complexity and deep uncertainty across spatiotemporal scales. Despite growing interest, few studies systematically integrate both concepts when assessing resilience. This paper conducts an integrative review of 102 English-language publications to identify gaps in current approaches. Findings reveal that most papers address lower levels of uncertainty and rarely consider dynamic complexity and deep uncertainty simultaneously, which limits the effectiveness of resilience strategies. To advance systems research, we propose a conceptual framework and practical tools to support researchers and decision-makers in evaluating and improving resilience. The paper also outlines future research directions for more robust, adaptive, and integrative resilience assessments.

**Funding**: This work was supported by the Dutch Research Council (NWO) [grant number COMPL.21COV.001].

# Resilience of coupled systems under deep uncertainty and dynamic complexity: An integrative literature review

1. **Introduction**

Recent global hazards - such as climate change and the COVID-19 pandemic - have intensified the need to assess and enhance the resilience of systems, particularly coupled systems (e.g., Berrouet, Machado, & Villegas-Palacio, 2018; Chua et al., 2020; Wernli et al., 2021). These refer to interconnected and interdependent systems that interact with each other. Disruptions in one component can cascade across others, amplifying systemic risk. Coupled systems can be found in various domains, including technological, ecological, social, and economic contexts. Examples include socio-technical systems, infrastructure networks, and socio-ecological systems, or more generally, systems of systems. Understanding and strengthening the resilience of such systems is essential for mitigating the impacts of complex, evolving threats.

Although there are many definitions of system resilience, the underlying idea of this integrative review is that regardless of the perspective of resilience considered, the amplitude of the largest shock that systems can stand without losing its long-term functionality (Béné & Doyen, 2018, p. 980). This also holds for resilience of coupled systems. Recent studies emphasise the need for a deeper understanding of two critical dimensions in resilience research: deep uncertainty and dynamic complexity (Biggs et al., 2012; Hogeboom et al., 2021; Ivanov & Dolgui, 2022). These dimensions are particularly relevant in the context of coupled socio-technical systems, where interactions between human actors, technological components, and environmental factors lead to emergent behaviour that is difficult to predict or control.

An example is the smart grid - a modern electricity infrastructure that integrates digital technologies to manage generation, distribution, and consumption. Unlike physical threats, which can often be analysed using historical data, cyberattacks pose substantial uncertainties. The timing, nature, and method are inherently unpredictable, complicating risk assessment and mitigation strategies. Moreover, smart grids involve a heterogeneous set of stakeholders, including utility companies, consumers, and regulatory agencies, all interacting through advanced technologies such as sensors, automated control systems, and real-time data analytics. A targeted cyberattack - such as malware injection into a control system - can trigger a series of adverse effects. It can disrupt power supply, impair communication networks, and jeopardise data integrity, ultimately resulting in widespread outages and a loss of consumer trust. Consequently, advancing resilience of coupled systems necessitates a robust conceptual framework that explicitly accounts for deep uncertainty and dynamic complexity.

Deep uncertainty differs from traditional uncertainty and risk analysis, which typically is based on probability distributions (e.g., W. E. Walker et al., 2003). Dynamic complexity refers to changes in system behaviour over time, such as exponential growth or decay, chaos, and oscillation, which emerge from causal interactions and feedback loops among activities (i.e., variables) within and between systems and their environment, across different spatiotemporal scales (Holland, 2006; Levin et al., 2013). The main reason deep uncertainty and dynamic complexity are separately considered in the current paper is that the latter does not explicitly take the former into account.

In this study, we conduct an integrative literature review (Snyder, 2019) to answer the following research question: "What constitutes a robust methodology for assessing and improving resilience in coupled systems under conditions of deep uncertainty and dynamic complexity?" The aim is to uncover a conceptual framework and tools for comprehensively examining the resilience of coupled systems amidst deep uncertainty and dynamic complexity. Focusing explicitly on coupled systems is essential due to the intricate interactions and feedback mechanisms between systems. These distinctive dynamics lead to emergent properties and nonlinear behaviours that are not present in single systems. Understanding these complexities is crucial for understand effective resilience strategies, as ignoring them may result in oversimplified solutions that fail to address the interconnected nature of coupled systems (Liu et al., 2021).

This paper is structured as follows. In Section 2, we present a causal framework in which the concepts 'deep uncertainty', 'dynamic complexity', and 'coupled system resilience' are explained and interconnected. We use the causal framework to define the criteria for the review of the selected papers. Section 3 explains the methodology applied in the integrative review, while the findings are described in Section 4. The results of the analysis of current gaps and research options to close these gaps are presented and discussed in Section 5, and the conclusion and limitations follow in Section 6.

## 2. Theoretical framework

This section provides an in-depth definition of the concepts 'deep uncertainty', 'dynamic complexity', and 'resilience assessment and improvement'. To connect these three concepts, a framework is chosen that is often used for understanding coupled complex systems, i.e., the 'Driver – Pressure – State – Impact – Response' (DPSIR) framework (EEA, 1999). The original DPSIR framework is deterministic, meaning it does not include uncertainty, let alone deep uncertainty. The deterministic approach of the DPSIR framework inevitably downplays the uncertainty and complexity of environmental events or hazards in light of coupled systems functioning. We will show that the five components of this framework can be used to understand the causal interactions and feedback loops between the environment and coupled systems and the responses of decision makers. Furthermore, the framework provides a solid foundation for supporting the improvement of the resilience of coupled systems (Zhao, Fang, Liu, & Liu, 2021).

## 2.1 Deep uncertainty and the Driver-Pressure-State-Impact-Response (DPSIR) framework

Coupled systems typically involve multiple decision makers who have varied and sometimes conflicting interests regarding the functioning and performance of the (sub)system in which they are involved. Disagreements among decision makers are likely to persist or intensify, especially when the coupled system is subjected to hazards such as those posed by global climate change and pandemics like COVID-19. For decision makers, it can be exceedingly challenging to ascertain the long-term effects of such hazards on the coupled system across various spatial scales. This phenomenon, characterised by 'known unknowns' is termed deep uncertainty. As Marchau, Walker, Bloemen, and Popper (2019, p. 2) point out, this situation implies "the experts do not know or the parties to a decision cannot agree upon (i) the external context of the system, (ii) how the system works and its boundaries, and/or (iii) the outcomes of interest form the system and/ or their relative importance." To illustrate deep uncertainty within a specific coupled system confronted by a distinct hazard, consider a Riverland Basin, in which water resource management is of paramount importance (Loucks & Van Beek, 2017). A basin accommodates various activities, ranging from agriculture practices to urban development (i.e., human system activities), while sustaining a diverse ecosystem (i.e., natural system activities). Imagine that in one particular year, the basin endures an unprecedented drought, leading to a significant decline in river water levels that affects all users. Farmers struggle to irrigate their crops, urban regions face water restrictions, and the ecosystem suffers from reduced water flow. The unexpected intensity of the drought underscores the challenges associated with forecasting climate-related impacts. Additionally, future climate projections indicating a rise in both the frequency and severity of droughts further increase the uncertainty.

Deep uncertainty arises from various sources associated with the five components of the DPSIR framework (Figure 1). The first source is *contextual* uncertainty, linked to the 'Driver' and 'Pressure'. For instance, the precise timing, magnitude, and severity of a potential future pandemic (Driver) remain uncertain. Furthermore, individual responses (Pressure) to governmental measures during a pandemic can be unpredictable as well. The second source of deep uncertainty pertains to uncertainties regarding the *structure and functioning* of systems, which relates to the State component of coupled systems. Although the structure and functioning of a single system or coupled systems may be known, the influences of drivers and pressures can induce structural changes that are challenging to forecast. The third source is *parameter* uncertainty, which refers to uncertainties in the specification and calibration of parameters within (future) system models. This source of uncertainty affects all components of the DPSIR framework. The fourth source of deep uncertainty is the prediction error associated with system model *outcomes*, which is tied to the 'Impact' component. Finally, the fifth source of deep uncertainty emphasizes the *relative importance* decision makers assign to specific outcomes and relates to the 'Response' component.

## 2.2 Dynamic complexity and Driver-Pressure-State-Impact-Response (DPSIR) framework

Dynamic complexity refers to the intricate behaviour of systems resulting from *multi- and cross-scale causal interactions* and *feedback loops* between exogenous and endogenous variables, affecting system functioning (Li, Dong, & Liu, 2020). Exogenous variables are independent variables typically found in the Driver and Pressure of the DPSIR framework, while endogenous variables are usually variables located in the State component. The terms 'multi scale' and 'cross scale' indicate interactions occurring across temporal (short to long term) and spatial scales (local to global). A positive causal interaction occurs when a change in one variable affects other variables in the same direction, whereas a negative causal interactions leads to changes in the opposite direction (Sterman, 2002). For example, a negative causal interaction observed in the context of COVID-19 in the United States is that states implementing mandatory face mask measures experienced a decrease in the cumulative number of infections (Chernozhukov, Kasahara, & Schrimpf, 2021). Feedback loops can be balancing or reinforcing, with balancing loops seeking to stabilise changes and reinforcing loops amplifying initial changes (Sterman, 2002), potentially leading to periods of chaos and counterintuitive behaviour in systems (i.e., nonlinearity). Both causal interactions and feedback loops occur within and among the Driver, Pressures, State, and Response components (Figure 1).

## 2.3 Resilience assessment and improvement of coupled systems under deep uncertainty and dynamic complexity integrated into the Driver-Pressure-State-Impact-Response (DPSIR) framework

The resilience literature encompasses four different viewpoints for assessing and improving resilience: engineering resilience, resilience engineering, ecological resilience, and evolutionary resilience. Engineering resilience views a system as being near to stable steady-state, focusing on the system's capability to return to this state quickly after a disturbance (e.g., Holling, 1996). Resilience engineering concentrates on (often socio-technical) systems managing complexity and embracing performance variability (Patriarca, Bergström, Di Gravio, & Costantino, 2018). Similar to engineering resilience, resilience engineering strives to address disturbances so that systems can promptly resume normal operations with minimal performance decline (Fairbanks et al., 2014). Ecological resilience, in contrast, operates far from stable steady-states, emphasizing a system's ability to endure disturbances and uphold essential functions (e.g., Adger et al., 2011; van der Merwe, Biggs, & Preiser, 2018). Lastly, evolutionary resilience is closely linked to adaptation, and focus on a system's capacity to adapt and endure over time, even in changing circumstances (Davoudi et al., 2012). Initial examination of the various resilience viewpoints indicates that ecological and evolutionary resilience emphasize deep uncertainty and dynamic complexity to a greater degree. Resilience engineering, akin to engineering resilience, places a central emphasis on (dynamic) complexity. Nevertheless, the level of integration of the sources of deep uncertainty and dynamic complexity indicators within the diverse resilience perspectives remains unclear.

Looking at the DPSIR framework (Figure 1), the assessment of resilience can be situated within the Impact component, while the improvement of resilience can be situated within the Response component. To assess and improve resilience under conditions of deep uncertainty and dynamic complexity, it is essential to understand the key constructs and indicators used in resilience evaluation across different perspectives, and how these relate to deep uncertainty and dynamic complexity.

In the original DPSIR framework, there are causal interactions between Response → Driver and Response → Pressure (EEA, 1999). However, particularly in the context of complex coupled systems, these causal interactions are debatable and will not be incorporated into the framework. It can be argued that the actions (Response) taken by the involved decision makers of a coupled system (e.g., decision makers in healthcare and education) do not have a direct effect on the Driver (e.g., a pandemic) nor on the Pressure (e.g., human behaviour in response to the pandemic), but rather an indirect effect, i.e., through the system State and Impact. Additionally, the spatial scale also plays a key role. For example, when the responses occur at a local or regional scale, but the drivers and pressures are on a national or international scale, then the effectiveness of such responses on Driver and Pressure may be limited, at least initially. Moreover, the existing literature does not provide insight into the tools that can be used to analyse the causal interactions within the DPSIR framework, especially from the standpoint of coupled system resilience. The next section describes the methodology applied in the integrative literature review.

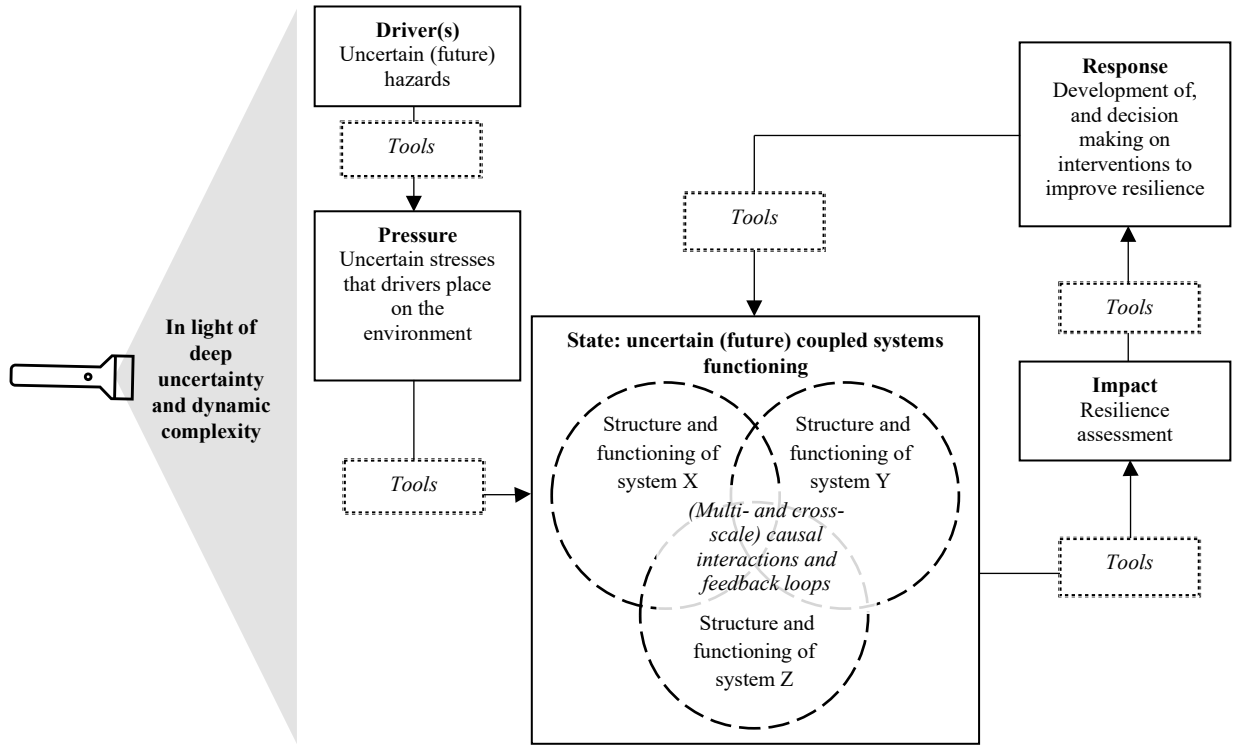

**Figure 1.** Proposed Driver-Pressure-State-Impact-Response (DPSIR) framework for resilience assessment and improvement of coupled systems under deep uncertainty and dynamic complexity.

## 3. Research methodology

An integrative literature review was conducted by incorporating the three key concepts (deep uncertainty, dynamic complexity, and resilience assessment and improvement) along with the four resilience perspectives. The aim of an integrative literature review is to study the existing literature on a specific research phenomenon in order to uncover new theoretical and conceptual insights (Snyder, 2019). Given the absence of a rigid protocol for data collection and analysis in integrative literature reviews, we followed the systematic approach outlined by Tranfield, Denyer, and Smart (2003) as depicted in Figure 2. Step 1 involved breaking down the research question outlined in Section 1 into two distinct sub-questions that guided the literature search:

1. How is the assessment and improvement of the different resilience perspectives under deep uncertainty and dynamic complexity operationalised?
2. What tools have been applied to assess and improve resilience of coupled systems in the presence of deep uncertainty and dynamic complexity?

Subsequently, a comprehensive set of search strings, as detailed in Table 1, was utilized to explore the databases of Web of Science, ScienceDirect, and IEEE Xplore in July 2024. These keywords encompassed the terms resilience assessment or resilience improvement, and coupled systems. Moreover, the focus was on identifying papers on system resilience that explicitly dealt with deep uncertainty or dynamic complexity. The search and selection phase did not confine itself solely to the overarching term coupled systems; it also encompassed related keywords like socio-ecological system, socio-technical system, and system of systems. This comprehensive search yielded a total of 442 hits. Subsequently, papers in English were selected and duplicates were eliminated, resulting in a set of 281 papers. These papers were further scrutinised based on their title and abstract to identify those specifically addressing resilience in the context of deep uncertainty or dynamic complexity, leading to a selection of 50 papers. To ensure the inclusivity of relevant literature, a snowballing approach (both forward and reverse) was employed (Wohlin, 2014), starting from the initial set of 50 papers. Forward snowballing involves identifying new papers by examining those that cited one or more of the 50 papers under review. Backward snowballing entails exploring the reference lists of the 50 papers to identify additional relevant papers. The snowballing procedure resulted in the inclusion of 52 additional papers, bringing the total number of papers selected for in-depth analysis to 102.

Two analyses were conducted: a conceptual analysis and an analysis of the tools utilized in the selected papers. Categorisation formed the basis for conducting these analyses. Prior to data collection, we established categories for the conceptual analysis, which are detailed in Figure 3 and further illustrated in the supplementary material. As for the analysis of the tools applied, categories were developed a posteriori, based on the research methods employed, such as a participatory approach or a scenario planning approach.

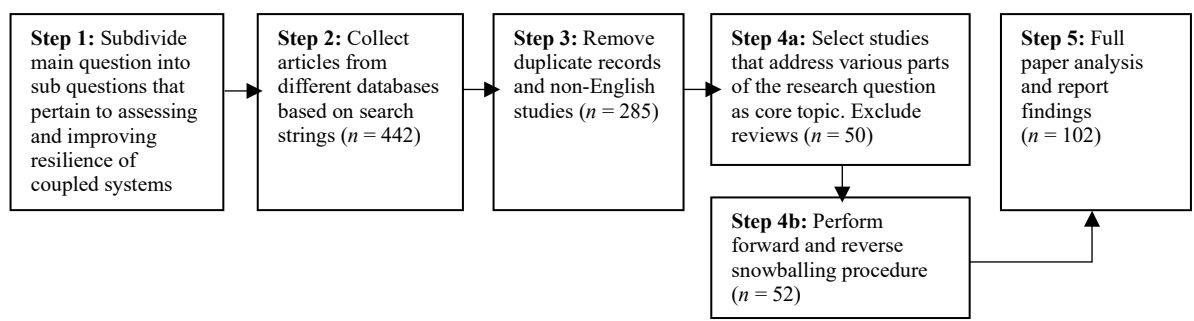

**Figure 2.** Step-by-step procedure for integrative review.

**Table 1.** Search strings and resulting numbers of hits.

| Search strings/keywords | Total number of search strings used |
|---|---|
| deep uncertainty AND resilience assessment OR resilience improvement AND coupled systems OR linked systems OR socio-ecological system OR socio-technical system OR system of systems* | 10 |
| dynamic OR complexity AND resilience assessment OR resilience improvement AND coupled systems OR linked systems OR socio-ecological system OR socio-technical system OR system of systems | 20 |
| **Total** | **30** |

* For instance: deep uncertainty AND resilience assessment AND coupled systems

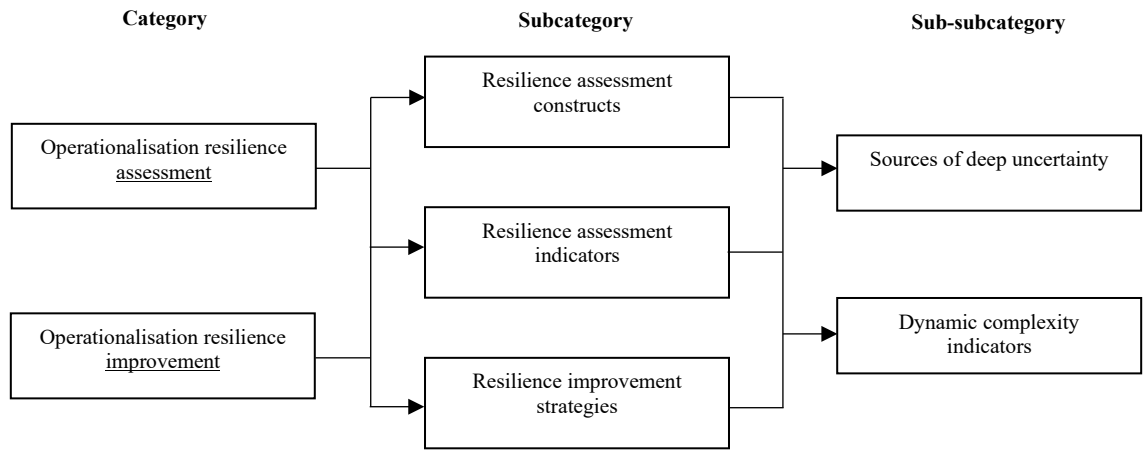

**Figure 3.** A-priori categorisation of conceptual analysis.

## 4. Results

This section presents the conceptual findings and the tools employed in the analysed literature. The conceptual results were focused on addressing Sub-Question 1, while the tools utilized in the analysed papers aimed at addressing Sub-Question 2.

### 4.1 Conceptual findings

The analysis indicated that the key constructs of assessing resilience in coupled systems encompass robustness, recovery rapidity, *changes in the stability of coupled systems*, *connectedness*, *centrality*, *functional redundancy*, and *response diversity* (Figure 4). The initial trio of constructs are system-based, while the latter quartet is network-based. In general, deep uncertainty plays a role – albeit a minor one – in studies focused on system-based resilience assessment. Conversely, dynamic complexity has received more emphasis. A more detailed explanation of the constructs associated with deep uncertainty and dynamic complexity follows below.

**System-based resilience assessment constructs and indicators**

All of the examined studies focusing on the construct *recovery rapidity* (e.g., An et al., 2023; Yodo et al., 2017) have overlooked deep uncertainty, while the construct *robustness* is incorporated. For instance, research conducted by Little et al. (2019) and Freeman et al. (2020) examine the impact of unpredictable demographic changes and global climate change on the ability of coupled systems to withstand disturbances over time. Nevertheless, the majority of the studies tend to concentrate on predictable *changes in the stability of coupled systems*, emphasizing indicators such as the relationship between *slow* and *fast* variables (B. H. Walker, Carpenter, Rockstrom, Crépin, & Peterson, 2012), *latitude, precariousness,* and *resistance* associated with the so-called basins of attraction (e.g., J. Carper et al., 2021; Herrera, 2017; Shi et al., 2021).

Slow variables, exemplified by soil organic matter within natural systems, shape the responses of fast variables, such as crop production (economic system), to disturbances arising from hazards, such as seasonal rainfall. These interactions can be linear and nonlinear, classifying systems into various basins of attractions, which can be either desirable or undesirable (Carpenter, Walker, Anderies, & Abel, 2001). The resilience of these basins is determined by latitude (the extent of change a basin of attraction can tolerate), resistance (ease or difficulty of altering basins), and precariousness (closeness to a critical threshold) (e.g., Kinzig et al., 2006). Notably, resistance is intricately connected to the resilience construct robustness.

Despite the relevance of precariousness in the context of deep uncertainty, this relationship is frequently overlooked in the resilience literature. For instance, the health of soil's may be precariously close to a critical threshold, where even a minor reduction in organic matter (slow variable) could severely impair its capacity to sustain crops (fast variable). This threshold is affected by numerous factors, including erosion, nutrient depletion, and extreme weather events. Given the deep uncertainties associated with future climate scenarios, predicting the effects of changes in rainfall and temperature on soil health becomes increasingly challenging. Should the soil surpass the threshold, both the natural and economic system could shift to an undesirable state that is difficult to revert, necessitating substantial effort and resources to restore soil health and agricultural productivity.

Therefore, in the context of system-based resilience assessment amid deep uncertainty and dynamic complexity, it can be posited that attention should be directed towards all indicators related to the construct changes in the stability of coupled systems, including precariousness, where robustness is an integral aspect of the indicator resistance.

**Network-based coupled system resilience assessment constructs and indicators**.
Research examining network-based resilience constructs have also largely overlooked deep uncertainty. The primary emphasis of these studies is on both temporal and spatial scales, as well as the implications of system interconnectedness for resilience. For instance, the construct *connectedness* pertains to critical nodes and their interconnections across different scales (Carvalhaes, Chester, Reddy, & Allenby, 2021). A pertinent example is the dual role of high connectedness during the SARS-CoV-1 outbreak, which not only facilitated the rapid transmission of the virus, but also allowed for the quick sharing of information regarding treatments (Janssen et al., 2006). Connectedness is measured by *density*, defined as the number of links divided by to the maximum possible links, and *reachability* (accessibility of all nodes). Higher density and reachability indicate greater resilience within the coupled systems.

The construct *centrality* focuses on the individual nodes within coupled systems and is assessed by *degree centrality* (number of direct links to a specific node) and *betweenness centrality* (the role of nodes as intermediaries for resource transfer among other nodes) (e.g., Vercruysse, Deruyter, De Sutter, & Boelens, 2024). A high degree centrality contributes to redundancy, thereby enhancing resilience. For instance, imagine that a local community relies on multiple fish species (nodes) for their livelihood, in which fish species 1 is abundant and widely distributed (high degree centrality), while fish species 2 is less abundant. If fish species 2 were to decline, the community can still rely on fish species 1. This redundancy ensures the community is resilient, as the decline of one species does not substantially disrupt their livelihood. In contrast, betweenness centrality measures reflect a node's capacity to influence resource distribution, which can either positively or negatively affect resilience across various scales.

The construct *functional redundancy* refers to the presence of nodes that can replace one another, thereby performing similar roles within the operations of coupled system across defined temporal and/or spatial scales (e.g., B. Walker et al., 2006). This construct is assessed based on the *number* and *diversity of nodes* associated with a given node (Cabell & Oelofse, 2012; Oliveira et al., 2022). Nodes that exhibit functional redundancy may exhibit varying responses to disturbances, also known as the construct *response diversity* (Hodbod, Barreteau, Allen, & Magda, 2016; Paas et al., 2021), which can bolster the resilience of coupled systems across different spatiotemporal scales (Sundstrom, Angeler, Garmestani, Garcia, & Allen, 2014).

The four network-based resilience constructs exhibit strong interconnections, albeit with different focuses. The indicators density and reachability assess the overall connectivity of coupled systems, whereas degree and betweenness centrality highlight the key nodes and their respective functions within these systems. Functional redundancy pertains to the backup capabilities of nodes, while response diversity examines the diversity in responses of nodes to disturbances. By integrating these indicators, one can assess both the structural and functional aspects that affect the resilience of coupled systems. Nonetheless, it is crucial to account for deep uncertainty in this matter. For example, consider a city reliant on a single communication system, such as satellites. If this system were to fail due to unexpected technological issues (e.g., cyber-attacks) or unforeseen surge in demand, the lack of alternative communication systems (like 5G or fibre optics) would hinder the maintenance of critical services, including public safety within the social system. Therefore, exploring the effects of such sources of deep uncertainty on the network-based resilience of coupled system is of critical importance.

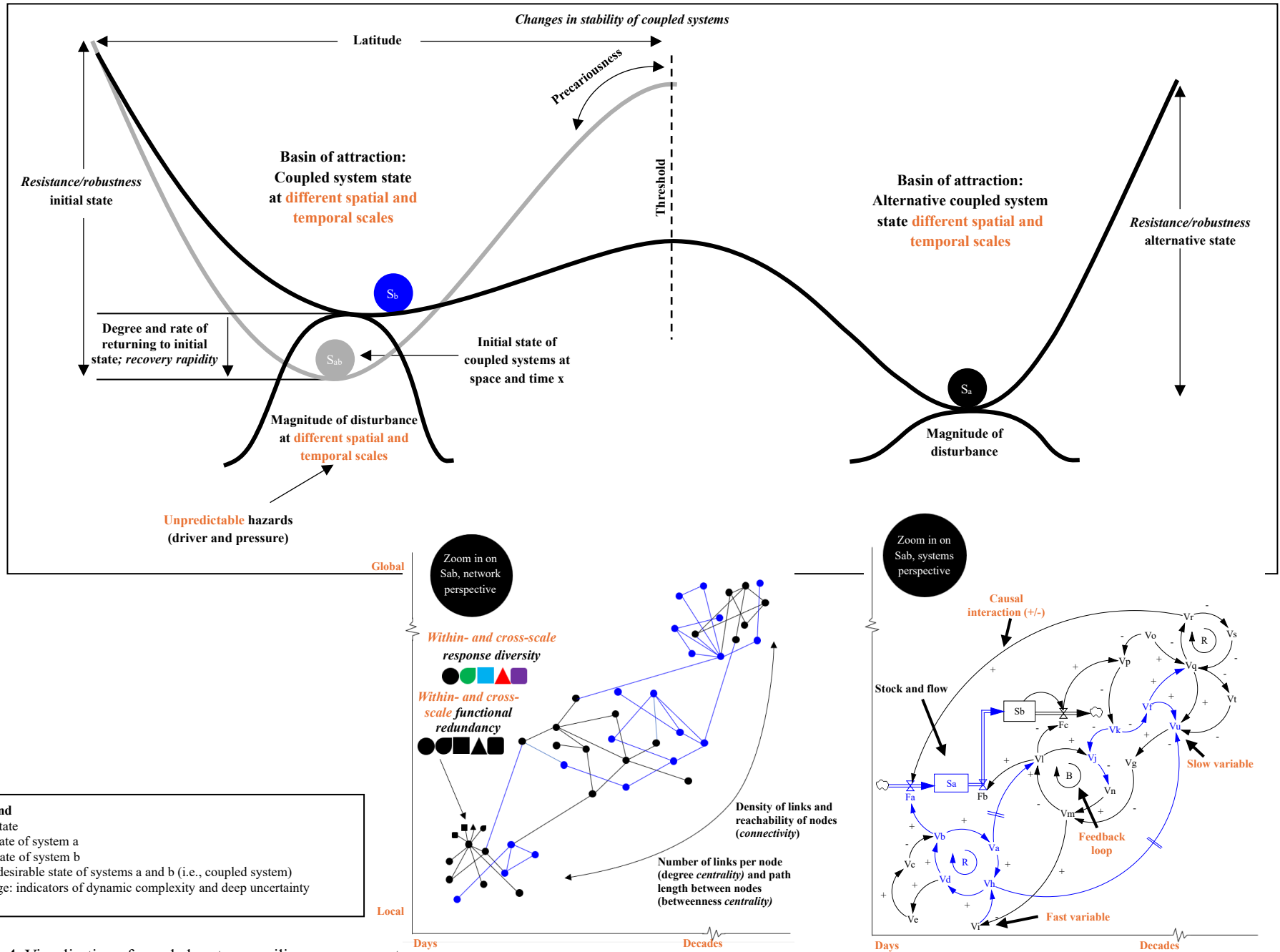

**Figure 4.** Visualisation of coupled systems resilience assessment.

Panarchy is a resilience assessment construct that involves adaptive cycles within a nested hierarchy, influencing each other across different scales of space and time (Peterson, 2000; Yu & Baroud, 2020). Each cycle has four phases: 'growth' (r), 'conservation' (k), 'collapse' (Ω), and 'reorganisation' (α) (Salvia & Quaranta, 2015) (see Figure 5). These cycles are interconnected, such as a local hospital's cycle affecting a network of regional hospitals, and vice versa. The two key indicators in panarchy are the *revolt function*, where changes in smaller cycles impact larger ones, and the *remember function*, which uses preserved potential for renewal (Berkes & Ross, 2016; Yongjun Yang et al., 2019).

Overall, panarchy incorporates minimal deep uncertainty but emphasizes dynamic complexity in terms of cross-scale interactions (e.g., Feofilovs & Romagnoli, 2020; Teigão dos Santos & Partidário, 2011) and feedback mechanisms, affecting both revolt and remember functions (Berkes & Ross, 2016; Yongjun Yang et al., 2019). In addition, one may argue that the revolt function can be linked to recovery rapidity, as rapid recovery in smaller cycles can influence larger cycles. The remember function can be associated with precariousness, as it involves using preserved potential to maintain coupled system stability.

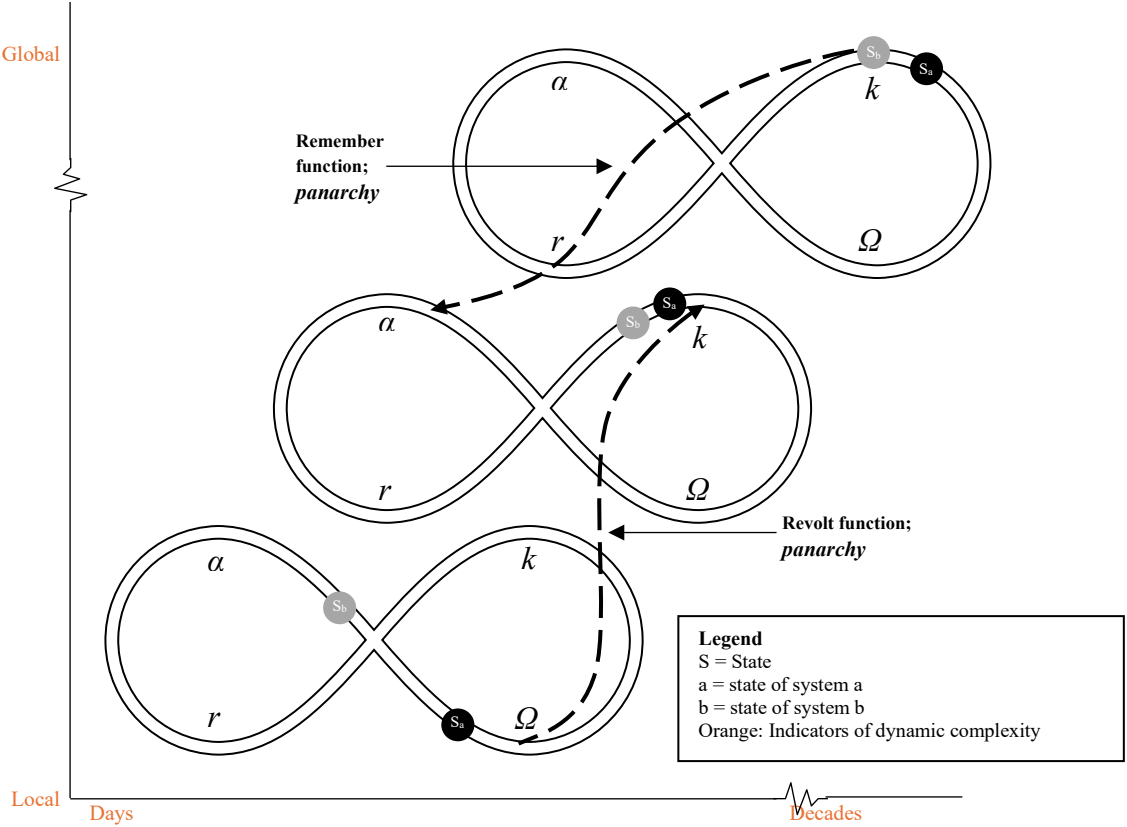

**Figure 5.** Visualisation of panarchy.

The resilience assessment construct *performance variability* is recognised as a contributor to both positive and negative outcomes within coupled systems (e.g., Aguilera et al., 2016; Jain, Pasman, Waldram, Pistikopoulos, & Mannan, 2018) (Figure 6). Key indicators for assessing performance variability include identifying *system functions and coupling*, *potential performance variability*, and *aggregation of variability* (e.g., Bellini et al., 2019; Bueno et al., 2021).

System functions involve activities necessary to achieve specific outcomes, and includes aspects like input, output, and resources (Patriarca, Di Gravio, & Costantino, 2017). System functions can be coupled upstream or downstream, depending on the sequence of aspects. The way functions are coupled influences how variability in one function affects others. Tight coupling can lead to rapid propagation of disturbances, impacting overall system stability. One may argue that this is like how changes in slow variables (like soil organic matter) affect fast variables (like crop production).

Potential performance variability involves analysing system functions under disturbance, their coupling with the external environment, and the responses of individual functions to performance variability (Bellini et al., 2019). High variability can lead to unpredictable outcomes, making a coupled system more precarious and thus closer to critical thresholds.

Aggregation of variability examines how variability in upstream and downstream functions interact. Nonlinear effects, such as resonance, can amplify disturbances, possibly leading to significant changes in coupled system stability (Stroeve & Everdij, 2017). One might argue that this can also affect the system's latitude (tolerance to change) and resistance (ease of altering states). For instance, aggregated variability can reduce the coupled system's latitude, making it less capable of tolerating changes, and lower its resistance, making it easier for disturbances to push one of multiple systems into a different state.

Although Stroeve and Everdij (2017) include deep uncertainty, in general, the concept is absent from the construct performance variability. Nevertheless, the construct is closely associated with deep uncertainty, as it is stated that ''variability makes the world inherently uncertain and unpredictable'' (Pruyt, 2007, p. 3). Furthermore, in relation to panarchy, one may argue that the phases of adaptive cycles can affect or are affected by performance variability of coupled systems. For example, during the growth phase, system functions may be more robust, while during collapse, variability may increase. In addition, precariousness can be assessed by examining potential performance variability and how close a coupled system is to critical thresholds. Latitude and resistance can be evaluated by analysing how variability in system functions affects a coupled system's capacity to tolerate changes and resist disturbances.

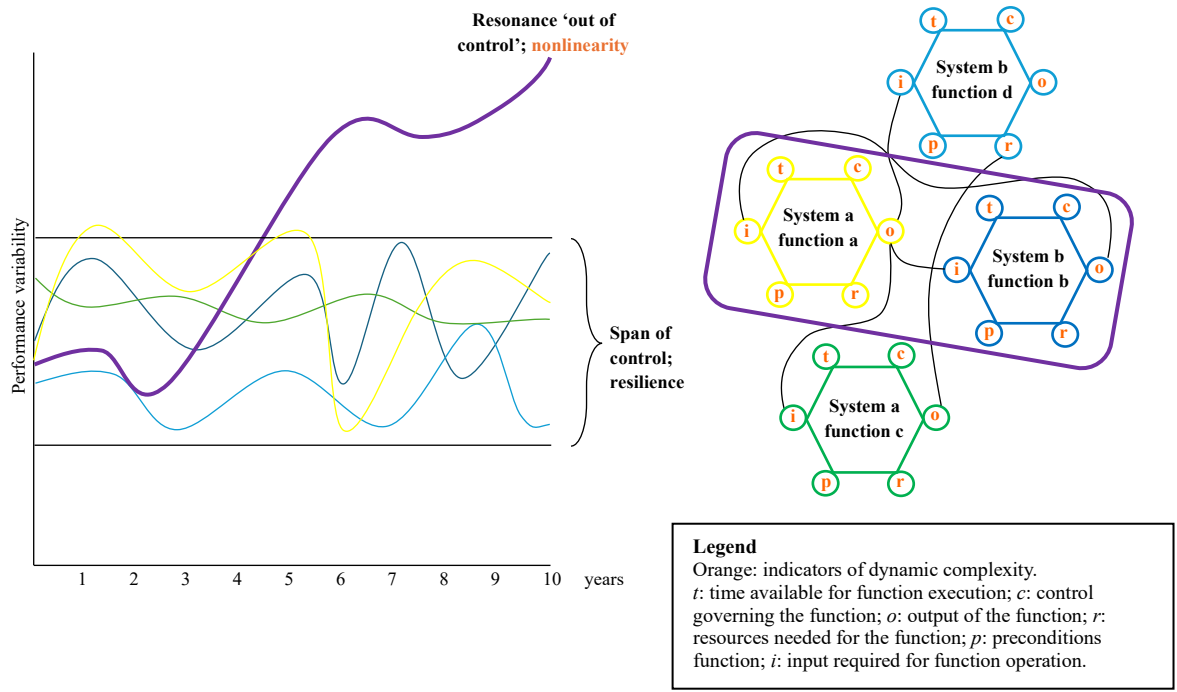

**Figure 6.** Visualisation of coupled system performance variability.

**Coupled system resilience improvement strategies**.

The resilience of coupled systems can be improved through a variety, and combination of strategies, including the fortification of system robustness, acceleration of recovery processes, provision of functional redundancy, fostering response diversity, and threshold management (e.g., Chen, Jayaprakash, & Irwin, 2012; Lu, Sun, & Steffen, 2023). For example, in the context of a healthcare-aquatic ecosystem: hospital physicians prescribe various medications, which patients excrete or dispose of improperly, leading to pharmaceuticals entering wastewater systems. These contaminants enter rivers, lakes, and oceans, affecting aquatic ecosystems by interfering with the endocrine systems of aquatic life and accumulating in organisms, posing health risks to humans. Polluted water sources can also compromise drinking water quality. To strengthen the resilience of a healthcare-aquatic ecosystem, it is important to enhance system robustness by using technologies that effectively remove drugs from wastewater. Continuous and thorough monitoring of antibiotic levels and resistant bacteria in water bodies is also essential to catch early signs of problems. Moreover, developing and promoting treatments that are less harmful to the environment is crucial. At the same time, removing drugs from wastewater requires energy and resources, necessitating a holistic approach that considers both the immediate and long-term societal, ecological, and economic effects.

Furthermore, alongside the previously discussed strategies, those that prioritise social learning, adaptive co-management, and the management of cross-scale interactions among adaptive cycles of coupled systems are also significant (e.g., B. Walker, Holling, Carpenter, & Kinzig, 2004; Yongjun Yang et al., 2019). In terms of managing performance variability, one effective approach is to amplify the positive aspects of such variability while mitigating its negative effects, which can be facilitated through improved resource allocation (Patriarca et al., 2017).

To strengthen the robustness of coupled systems, three studies have incorporated the concept of deep uncertainty (Dreesbeimdiek, von Behr, Brayne, & Clarkson, 2022; Rasoulkhani, Mostafavi, Reyes, & Batouli, 2020; Singh, Reed, & Keller, 2015). These studies focus on identifying trade-offs among anticipatory strategies and reassessing their robustness in deep uncertainty, thereby facilitating informed decision making for resilience management. Studies that incorporate dynamic complexity, focus on resilience-enhancing strategies such as co-management and threshold management. For example, co-management seeks to modify the latitude, resistance, and precariousness of basins of attraction through threshold management, which entails adjusting thresholds to either distance from or to approach the current states of coupled systems (Cutter et al., 2008; Olsson, Folke, & Berkes, 2004; B. Walker et al., 2004). Consequently, the emphasis lies on reinforcing existing balancing feedback loops, creating new feedback loops, and sometimes destabilising current loops within and among coupled systems (Enfors-Kautsky, Järnberg, Quinlan, & Ryan, 2021).

## 4.2 Tools used in extracted papers

The analysis revealed that a range of tools are used for assessing and improving the resilience of coupled systems facing deep uncertainty or dynamic complexity (see Table 3 for a detailed account). Based on the analysis, four distinct categories of tools have been identified: Modelling tools, scenario planning tools, analytical and optimisation tools, and participatory tools.

**Modelling tools**

The analysis highlighted system dynamics modelling and the functional resonance analysis method as key tools for assessing and improving resilience. System dynamics modelling is applied to all resilience constructs, except for panarchy and performance variability. This modelling tool typically incorporates the construction of a causal loop diagram to explore causal interactions and feedback loops (e.g., Datola, Bottero, De Angelis, & Romagnoli, 2022; San Martín et al., 2021; Sendzimir et al., 2011). For example, Sendzimir et al. (2011) created a causal loop diagram to explore reinforcing feedback loops within a socio-ecological system that affect the resilience of regional food security and tree density. This diagram can be modified or extended into a stock-flow diagram for quantitative analysis, where stocks denote resource quantities, and flows indicate temporal changes, as shown by Simonovic and Peck (2013). Notably, the studies using system dynamics modelling did not apply it for cross-scale analysis.

The functional resonance analysis method (FRAM) evaluates performance variability of coupled systems by examining nonlinear dependencies and fluctuations of coupled systems functions (Bellini et al., 2019). The tool models how various aspects (input, output, resource, and time) are interrelated to perform activities like operations and services (Eljaoued, Yahia, & Saoud, 2020). However, current modelling tools like FRAM fail to address deep uncertainty. Both Merino-Benítez, Bojórquez-Tapia,

Miquelajauregui, and Batllori-Sampedro (2024) and Little et al. (2019) recognised 'exploratory modelling' as a valuable technique to fill this gap, using system dynamics or agent-based modelling to generate a wide array of scenarios (also known as scenario discovery) and assess the robustness of different policy alternatives. Additionally, existing tools do not analyse cross-scale interactions, which are crucial for understanding panarchy of coupled systems.

**Scenario planning tools**

The reviewed studies used three distinct scenario planning tools to examine resilience: scenario development, scenario analysis, and scenario tree generation. Scenario development entails the formulation of narratives to explore plausible futures, highlighting the interactions between external (drivers and pressures) and internal environments (coupled systems) (e.g., Folke et al., 2002; Peng, Zhen, & Huang, 2023). For instance, Herrera and Kopainsky (2020) employed 'graphs-over-time', derived from causal loop diagrams, to illustrate the behaviour of key variables over time. However, the concept of deep uncertainty has not been explicitly incorporated into scenario development.

Scenario analysis is predominantly applied to assess recovery rapidity and changes in the stability of coupled systems within a defined specific spatiotemporal context (e.g., Datola et al., 2019; Fang, Lu, Li, & Hong, 2023; Filippini & Silva, 2014). For example, Pagano et al. (2017) conducted a scenario analysis to study the resilience of a local drinking water supply system following an earthquake, taking into account the evolving technical, social, and economic factors over a period of 100 days with daily evaluations. Like scenario development, deep uncertainty was not addressed in scenario analysis.

Goldbeck et al. (2019) applied a scenario tree generation algorithm to assess the robustness and recovery rapidity of interdependent urban infrastructure systems, specifically focusing on the metro and electric power networks in London. Scenario trees typically represent various scenarios along with their associated probabilities (Heitsch & Römisch, 2009), yet they do not capture deep uncertainty.

**Analysis and optimisation tools**

The selected papers did not use tools to analyse and optimise panarchy. Nevertheless, it can be posited that discontinuity analysis may serve this purpose. Discontinuity analysis has been applied by various researchers to study other resilience constructs, such as changes in the stability of coupled systems and response diversity (e.g., Chuang et al., 2018; Sundstrom et al., 2014), and it could similarly be applied to the study of panarchy. Discontinuity analysis facilitates the identification of scales within a coupled system and assesses it resilience by scrutinising patterns and processes across different spatiotemporal scales, which is essential for understanding the adaptive cycles inherent in panarchy.

Simulation and sensitivity tools are frequently used to study the various resilience constructs, encompassing numerical simulations (e.g., Goldbeck et al., 2019; Kong et al., 2019b; Krueger et al., 2019), (Hamiltonian), Monte Carlo simulations (Eljaoued et al., 2020; Kong et al., 2019a; Yabe, Rao, & Ukkusuri, 2021), discrete event simulations (J. Carper et al., 2021), and multi-agent simulations (Rasoulkhani et al., 2020). The latter tool recognises the existence of deep uncertainty. For instance, Rasoulkhani et al. (2020) conducted multi-agent simulations to explore, rather than predict, the long-term resilience of water supply infrastructure among various deeply uncertain sea-level rise scenarios and diverse adaptation strategies, including risk preferences, investment levels, and decision intervals. Consequently, in the realm of complex coupled systems, deep uncertainty can substantially affect the sensitivity of model parameters and the resilience of these systems (Gao et al., 2016).

An optimisation tool that addresses deep uncertainty and has been applied in several papers is multi-objective robust optimisation (Freeman et al., 2020; Singh et al., 2015). For example, Singh et al. (2015) utilised multi-objective robust optimisation to achieve a balance between lake health and economic benefits under pollution control strategies, considering parameter uncertainty.

**Participatory tools**

A range of participatory tools were applied to facilitate the assessment and improvement of resilience, including focus groups, brainstorming, group model building, participatory action research, and the cup-and-ball diagram. Focus groups and brainstorming were instrumental in developing causal loop diagrams and facilitating multi-criteria decision-making processes related to urban resilience (Bottero et al., 2020). Cavallo and Ireland (2014) employed focus groups to increase awareness of the implications of complexity, such as cross-scale interactions, as well as deep uncertainty, which includes unpredictable contextual risks, thereby fostering learning in the development of resilience.

Group model building is applied to improve stakeholders' comprehension of causal interactions, feedback loops, delays, and system behaviour over time, employing workshops and tools such as graphs-over-time (e.g., Herrera & Kopainsky, 2020; San Martín et al., 2021). However, deep uncertainty was not integrated into the group model building process.

Participatory action research was implemented by Salvia and Quaranta (2015) to promote collaboration between researchers and participants, aiming to deepening the understanding of the resilience of coupled agriculture systems (durum wheat and livestock) from societal, economic, and ecological viewpoints. The cup-and-ball diagram (e.g., Figure 4) serves as a tool to assist stakeholders understand transitions and contemplate adaptive strategies for enhancing resilience, illustrated by the movement of the ball between different basins of attraction (Yongjun Yang et al., 2019).

**Table 3.** Tools applied by selected papers

| Tool and authors | A1 | A2 | B1 | B2 | C1 | C2 | D1 | D2 | E1 | E2 | F1 | F2 | G1 | G2 | H1 | H2 | I2 | Deep uncertainty included | Dynamic complexity included |
|---|---|---|---|---|---|---|---|---|---|---|---|---|---|---|---|---|---|---|---|
| *Modelling* | | | | | | | | | | | | | | | | | | | |
| Physics-based modelling (Yang, Yifan et al., 2019) | X | | | | | | | | | | X | | | | | | | No, focus on probabilities | Multiple scales |
| Network modelling with geographical information system (Kong et al., 2019a) | X | | X | | | | | | | | | | | | | | | No, focus on probabilities | Multiple scales; nonlinearity |
| Multi-layered network modelling (Kong et al., 2019b) | X | | | | | | D1 | | | | | | | | | | | No, deterministic | Multiple scales |
| Dynamic network flow modelling (Goldbeck et al., 2019; Ouyang, 2017) | X | X | X | | | | | | | X | | | | | | | | No, focus on probabilities | Multiple scales; nonlinearity |
| Dynamic modelling (using damped harmonic oscillator equation and finite element modelling) (Cumming, 2011; Mahmoud et al., 2018) | X | | | | | | | | | | X | | | | | | | No, focus on probabilities | Multiple scales; nonlinearity |
| Dynamic Bayesian (belief) networks coupled with system dynamic modelling (Caetano et al., 2024; Franco-Gaviria et al., 2022; Pagano et al., 2017; Yodo et al., 2017) | | | X | | | | | | | | X | | | | | | | No, focus on probabilities | Causal interactions; multiple scales; nonlinearity; feedback loops |
| (Quantified) functional resonance analysis method (e.g., Aguilera et al., 2016; Bellini et al., 2019; Bueno et al., 2021; Steen et al., 2020) | X | | | | | | | | | | | X | | | X | | X | No, focus on probabilities or fuzzy logic | Nonlinearity |
| Agent-based modelling (Mostafavi et al., 2014; Pumpuni-Lenss et al., 2017; Stroeve et al., 2017) | | | X | X | | X | | | | | | | | | | | | No, focus on probabilities | Multiple scales; nonlinearity; causal interactions |
| Network modelling (Janssen et al., 2006) | | | | | X | | | | | | | | | | | | | No | Multiple scales |
| Physical-group built system dynamics modelling (SDM) (Carper et al., 2021; Carper et al., 2022) | X | X | X | X | X | X | X | | X | X | X | | | | | | | No, deterministic | Multiple scales; causal interactions; feedback loops; nonlinearity |
| (Group-built) SDM with or without causal loop and stock-flow diagramming (e.g., Bottero et al., 2020; Chuang et al., 2018; Herrera et al., 2020; Oliveira et al., 2022; Sendzimir et al., 2011) | X | X | X | X | X | X | X | | X | X | X | | | X | | | | No, deterministic or focus on probabilities | Multiple scales; causal interactions; nonlinearity; feedback loops |
| *Scenario planning* | | | | | | | | | | | | | | | | | | | |
| (Participatory) scenario development (e.g., Carper et al., 2022; Folke et al., 2002) | | X | X | X | | | | | | | X | | | | | | | Yes, focus on contextual uncertainty | Multiple scales; causal interactions; feedback loops |
| Scenario analysis (e.g., Datola et al., 2019; Pagano et al., 2017; Ross et al., 2021) | | | X | | | | | | | | X | | | | | | | No | Multiple scales; feedback loops; causal interactions; nonlinearity |
| Scenario tree generation (Goldbeck et al., 2019) | X | | X | | | | | | | | | | | | | | | No, focus on probabilities | Multiple scales |

Items registered with a 1 are resilience assessment constructs, and items registered with a 2 are resilience improvement constructs: A1: robustness; A2: improve robustness; B1: recovery rapidity; B2: improve robustness; adaptive/anticipative (co) management through threshold management; C1: connectivity; C2: improve recovery rapidity; D1: functional redundancy; D2: manage connectivity; E1: response diversity; E2: maintain redundancy and diversity; F1: changes in stability of coupled systems; F2: foster (social) learning; G1: panarchy; G2: manage slow variables; H1: performance variability; H2: manage cross-scale interactions; I2: manage performance variability

**Table 3.** (Continued).

| Tool and authors | A1 | A2 | B1 | B2 | C1 | C2 | D1 | D2 | E1 | E2 | F1 | F2 | G1 | G2 | H1 | H2 | I2 | Deep uncertainty included | Dynamic complexity included |
|---|---|---|---|---|---|---|---|---|---|---|---|---|---|---|---|---|---|---|---|
| *Analysis and optimisation* | | | | | | | | | | | | | | | | | | | |
| Sensitivity analysis (e.g., Chen et al., 2012; Homayounfar et al., 2022; Oliveira et al., 2022; Pagano et al., 2017; Suweis et al., 2015) | X | X | X | X | X | X | X | X | X | X | X | X | | | | | | No | Multiple scales; causal interactions; nonlinearity; feedback loops |
| (Multi-agent) simulations (e.g., (Hamiltonian) Monte Carlo) (e.g., Janssen et al., 1999; Watson et al., 2021) | X | X | X | X | X | X | X | X | X | X | X | X | X | X | X | | X | Yes (Rasoulkhani et al., 2020) | Multiple scales; feedback loops; causal interactions; nonlinearity |
| Multi-scale/cross-scale analysis (e.g., discontinuity or fractal analysis) (Chuang et al., 2018; Zurlini, Riitters, et al., 2006; Zurlini, Zaccarelli, et al., 2006) | | | | | X | | X | | X | | X | | | | | | | No | Across multiple scales; feedback loops; causal interactions; nonlinearity |
| Multi-criteria decision aid tools (Bottero et al., 2020; Feofilovs et al., 2020) | | | | | | | | | | | X | | | | | | | No | Nonlinearity; causal interactions; feedback loops |
| Multi-objective robust optimisation (Freeman et al., 2020; Singh et al., 2015) | X | X | | X | | | | | X | | X | | | | | | | Yes, focus on contextual, parameter and in outcome | No (focus on temporal scale only) |
| Markov decision process (Govindan et al., 2019) | | | | | | | | | | | | X | | | | | | No, focus on probabilities | Multiple scales; feedback loops |
| *Participatory tools* | | | | | | | | | | | | | | | | | | | |
| Focus groups and brainstorming – in some cases as part of scenario development (e.g., Aguilera et al., 2016; Cavallo 2014; Freeman et al., 2020) | X | X | | | X | | | | | X | X | X | | | X | X | X | Yes, contextual, parameter and outcome | Multiple scales (cross-scale); causal interactions |
| Group model building (e.g., Datola et al., 2019; Herrera et al., 2020; Paas et al., 2021) | X | | X | | | | | X | | X | | | | | X | | X | No | Causal interactions; feedback loops |
| Participatory action research (Salvia et al., 2015) | | | | X | | | | | | | X | | | X | | | | No | Multiple scales (cross-scale); causal interactions |
| Cup-and-ball diagram (Yang, Yongim et al., ...) | | | X | | | | | | | | | | | | | | X | No | Multiple scales |

Items registered with a 1 are resilience assessment constructs, and items registered with a 2 are resilience improvement constructs: A1: robustness; A2: improve robustness; B1: recovery rapidity; B2: improve robustness; adaptive/anticipative (co) management through threshold management; C1: connectivity; C2: improve recovery rapidity; D1: functional redundancy; D2: manage connectivity; E1: response diversity; E2: maintain redundancy and diversity; F1: changes in stability of coupled systems; F2: foster (social) learning; G1: panarchy; G2: manage slow variables; H1: performance variability; H2: manage cross-scale interactions; I2: manage performance variability

## 5. Discussion and new insights

This section proposes reflections on the review outcomes, emphasizing new and adjusted theoretical and conceptual insights with the aim of forming a solid framework to assess and improve coupled system resilience in light of deep uncertainty and dynamic complexity.

**Reconsidering the engineering resilience perspective**

The literature reviewed demonstrates a limited emphasis on deep uncertainty concerning both the assessment and improvement of resilience. Most of the studies that do consider deep uncertainty are rooted in ecological and evolutionary resilience perspectives. In contrast, the field of engineering resilience often prioritises stochastic distributions or employs a deterministic approach. Future research could explore the validity of engineering resilience when analysed under conditions of deep uncertainty. For example, a disruption in System A or a decision made regarding this system, may adversely impact the resilience of interconnected Systems B and C, in turn affecting System A itself (known as the rebound effect), resulting in deep uncertainty of initial system A predictions. In such cases, the strategy of quickly returning System A to its original equilibrium state may no longer be feasible. Therefore, a significant alteration in the operation of one or both systems is essential to sustain or enhance the resilience of the coupled systems, potentially through a (quasi) experimental approach.

**An integrated focus on deep uncertainty within resilience research of coupled systems**

The limited studies that have incorporated deep uncertainty primarily concentrated on contextual uncertainty, which corresponds to the Driver and Pressure components of the DPSIR framework. This focus is justifiable within the resilience discourse, as it often pertains to disturbances arising from external threats, including pandemics or natural disasters. Nonetheless, it is essential to acknowledge that deep uncertainty can also originate from the internal dynamics of coupled systems, reflected in the parameters and behaviours of these systems. For example, short-term decisions made by policymakers in System A may advertently result in long-term repercussions that could adversely affect the overall resilience of the system. Consequently, it is vital for resilience research to prioritise the exploration of deep uncertainty. Trying to predict how internal and external hazards of coupled systems will affect the resilience across various spatiotemporal scales may be overly optimistic, especially when considering complex threats like future pandemics, climate change, market shifts, or evolving decision-maker behaviour.

**Integrating network-based and system-based resilience constructs for resilience study under deep uncertainty and dynamic complexity**

A limited number of studies have employed both network-based and system-based resilience constructs to analyse the resilience of coupled systems (Datola et al., 2019; Oliveira et al., 2022; Zhang, Liu, Shi, Huang, & Huang, 2022). Nonetheless, these studies have focused on evaluating the constructs separately or performing comparative assessments. In the realm of coupled systems, such as socio-ecological and socio-technical systems, it is posited that the interaction between network-based and system-based resilience constructs is vital for understanding and improving system resilience, particularly under conditions characterised by deep uncertainty and dynamic complexity. For instance, a slight change in fishing efforts (fast economic variable) can lead to large, unpredictable changes in the fish population (slow ecological variable) due to feedback loops. If the fish population is already near a critical threshold (precariousness) due to various exogenous drivers and pressures, such as climate change and rising ocean temperatures, even a slight rise in fishing efforts could lead to a system collapse. However, functional redundancy may provide a buffer, as one species may offset the decline of others. Additionally, elevated market prices (fast variable) may encourage increased fishing efforts, further stressing the fish population. Response diversity can enable the system to absorb these pressures, as various species exhibit different reactions to heightened fishing efforts. It is proposed that the DPSIR framework facilitates an understanding of the complex interactions among the various components from both a network and system perspective (Figure 7), clarifying the role of each variable within the specific context. This approach aids researchers and stakeholders in studying and managing the resilience of complex coupled systems in deep uncertainty.

**Tools for studying the causal interactions between DPSIR components under deep uncertainty and dynamic complexity for resilience studies**

The tools used in the selected studies addressing deep uncertainty included participatory scenario development and multi-objective robust optimisation. Furthermore, 'exploratory modelling and analysis' was identified as a potential approach, although it was not implemented in the studies. This tool employs computational experiments to study complex systems characterised by deep uncertainty, with a focus on robustness (Kwakkel & Pruyt, 2013). Multi-objective robust optimisation is an integral part of this approach, alongside scenario discovery for vulnerability analysis and stress testing (Kwakkel, 2017). The strong emphasis on evaluating system robustness prompts inquiries regarding the relevance of this approach in assessing other resilience constructs, such as connectivity. Consequently, there is a need for further exploration into the suitability of various deep uncertainty tools for assessing different resilience constructs.

Additionally, the modelling tools used in the reviewed literature did not concurrently address deep uncertainty and dynamic complexity. Tools such as (group-built) system dynamics modelling and agent-based modelling possess the capability to tackle both aspects, providing a basis for assessing and improving resilience, particularly in terms of robustness (Kwakkel & Pruyt, 2013; Martin & Schlüter, 2015). Moreover, causal loop diagramming, which is a part of system dynamics modelling, can be combined with scenario development to explore plausible future scenarios, thereby enhancing research on coupled-system resilience (VandenBroeck, 2021).

Furthermore, various selected studies have conducted (local) sensitivity analysis within constrained parameter ranges. However, this tool fails to account for potential worst-case scenarios (Siegenfeld, Taleb, & Bar-Yam, 2020), and is insufficient for exploring the interactions of parameters that exhibit deep uncertainty in nonlinear systems (e.g., Saltelli et al., 2019). As a result, future research that acknowledges deep uncertainty and complexity may benefit from the implementation of a global sensitivity analysis. This analytical tool is designed to examine the effects of causal interactions among uncertain parameters and their implications for the outputs of simulation models (Gao et al., 2016). For example, in the realm of climate change, global sensitivity analysis could reveal how varying assumptions, such as the duration and intensity of droughts and soil composition, might converge to create a worst-case scenario for crop yields.

Because of deep uncertainty, decision makers often disagree about which target coupled system variables to focus on, which interventions will best improve resilience, and how important those interventions and their outcomes really are. Nevertheless, scholarly attention to how these challenges should be addressed has remained limited. In the literature on deep uncertainty, a process known as 'agree-on-decisions' is frequently used to foster consensus around robust decision making, and to aid in managing deep uncertainties and disagreements on these matters (Stanton & Roelich, 2021). Here, the primary objective is often not to identify the optimal solutions and achieve consensus at the outset, but to stress-test diverse options across a broad spectrum of plausible scenarios. The tools associated with agree-on-decisions include those previously mentioned, such as exploratory modelling, scenario discovery, as well as tools like multi-objective robust optimisation, real options analysis, and info-gap analysis (Kalra et al., 2014; Marchau et al., 2019). These tools are typically participatory, meaning that their implementation involves close collaboration with decision makers, experts, and stakeholders. Consequently, they can assist decision makers in evaluating interventions, formulating strategies, and clarifying preferences while explicitly taking into account the analysis of uncertainties (Stanton & Roelich, 2021).

The functioning of coupled complex systems takes places at distinct spatiotemporal scales, with interactions across various scales affecting scale-specific (nonlinear) responses, ranging from individual to the entire network, in reaction to disturbances. While several selected studies have conceptually addressed spatiotemporal scales (e.g., Hatt & Döring, 2023), the integration of appropriate tools to study the multi- and cross-scale effects remains a challenge, let alone under deep uncertainty. As noted, one potential tool for examining cross-scale patterns is discontinuity analysis; however, this tool does not account for deep uncertainty. Some efforts have been made to do so, such as the application of uncertainty quantification algorithms for multi-scale models. Nonetheless, uncertainty quantification frequently emphasizes the reduction of uncertainty rather than its acceptance. Conversely, tools designed to address deep uncertainty, such as info-gap analysis and scenario discovery, which also fall under the umbrella of uncertainty quantification, tend to overlook multi-scale modelling. Therefore, the deep uncertainty quantification of the multi-scale resilience of coupled systems warrants greater focus.

The tools used to examine interactions among the 'Driver-Pressure-State-Impact-Response' components under conditions of deep uncertainty and dynamic complexity have been integrated into the DPSIR framework, thereby significantly strengthening its analytical capacity (see Figure 7). We encourage future researchers and policymakers to further develop and apply this framework in the study and management of resilience in complex coupled systems across diverse contexts.

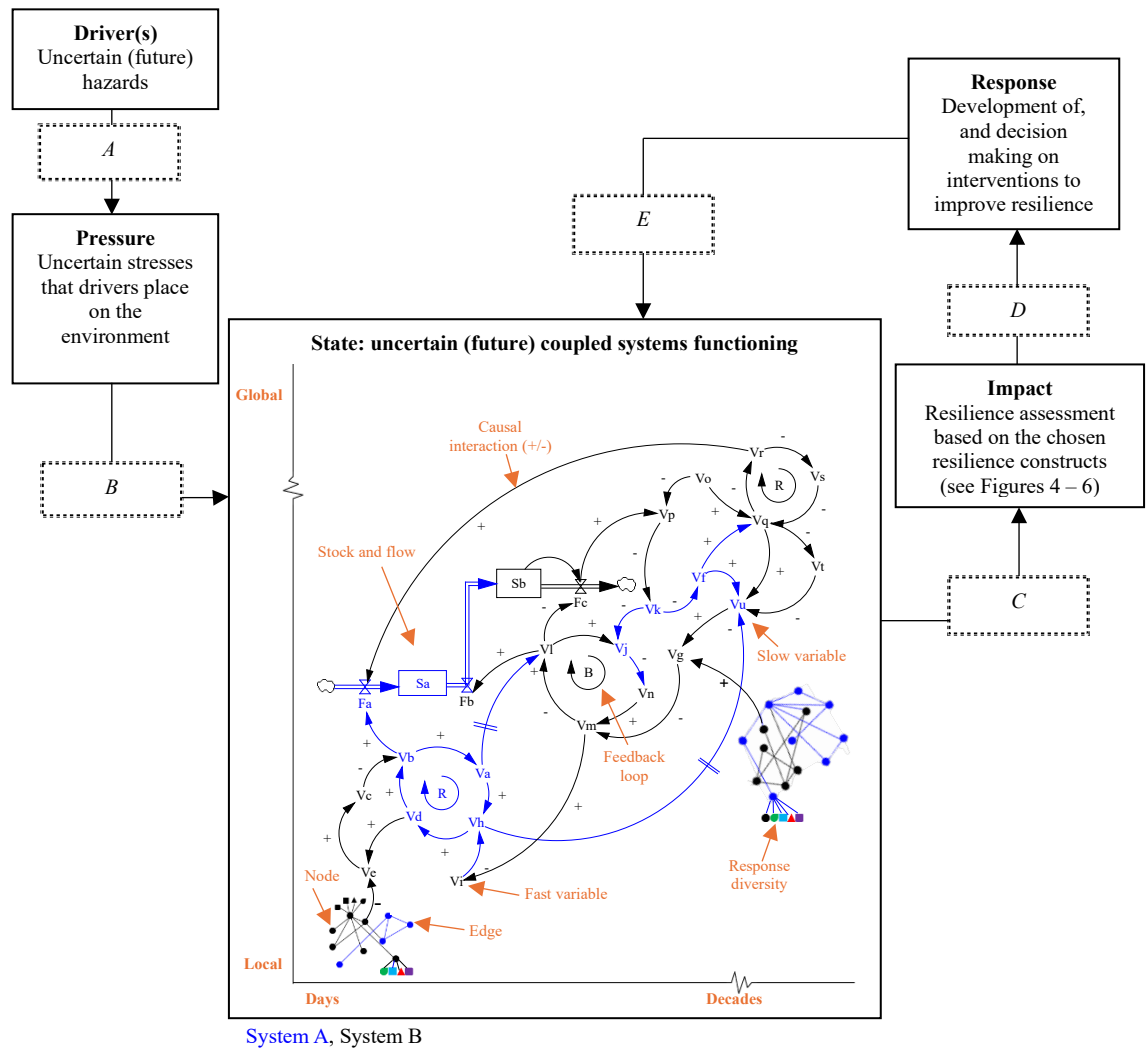

| Location of tools that acknowledge deep uncertainty or dynamic complexity, as a basis to connect deep uncertainty and dynamic complexity | |
|---|---|
| A | Participatory scenario development |
| B and E | Participatory scenario development – combined with – (group-built) exploratory system-dynamics or agent-based modelling; multi-scale modelling |
| C and D | Multi/many-objective robust optimisation; global sensitivity analysis; scenario discovery; discontinuity analysis; info-gap analysis; real options analysis |

**Figure 7.** An extension of the Driver-Pressure-State-Impact-Response (DPSIR) framework, grounded in literature, incorporating tools for assessing and improving resilience of coupled systems in a context of deep uncertainty and dynamic complexity.

## 6. Conclusion and limitations

This review was initiated by the inquiry into a solid approach for assessing and improving coupled system resilience among deep uncertainty and dynamic complexity as this proved to be lacking in the COVID-19 pandemic. Through an examination of selected literature on this topic, three main gaps were discerned: the insufficient recognition of deep uncertainty within coupled systems resilience research, a lack of adequate tools for analysing the causal interactions among DPSIR components in the context of deep uncertainty and dynamic complexity for resilience studies, and the absence of a synthesis between network-based and system-based resilience constructs for studies addressing these challenges. We aimed for closing these gaps through a motivated expansion of the DPSIR framework. Consequently, the study emphasises the necessity of studying resilience through the lens of coupled systems, with a particular focus on both deep uncertainty and dynamic complexity. The proposed causal framework is intended to assist this endeavour.

The review is constrained by two primary limitations. First, it focused solely on articles sourced from established databases, such as Web of Science. This approach may overlook other valuable publications, such as dissertations, which could provide insights not available in scientific journals. The decision to exclude dissertations stemmed from their scattered distribution online. Second, the literature review yielded a relatively modest number of hits, specifically 442. This outcome may be attributed to the chosen keywords and search strings, which, despite careful consideration, might not have captured the full scope of relevant literature, or it could indicate the restricted number of databases used. Expanding the search to include additional databases, like Scopus, could have yielded a greater number of results, albeit potentially increasing the incidence of duplicate entries. To improve the thoroughness of the studies included, the forward and reverse snowballing technique was used.

Oliveira, B. M., Boumans, R., Fath, B. D., Othoniel, B., Liu, W., & Harari, J. (2022). Prototype of social-ecological system's resilience analysis using a dynamic index. *Ecological Indicators, 141*, 1-9*.

Olsson, P., Folke, C., & Berkes, F. (2004). Adaptive comanagement for building resilience in social–ecological systems. *Environmental Management, 34*, 75-90*.

Ouyang, M. (2017). A mathematical framework to optimize resilience of interdependent critical infrastructure systems under spatially localized attacks. *European Journal of Operational Research, 262*(3), 1072-1084*.

Paas, W., Coopmans, I., Severini, S., Van Ittersum, M. K., Meuwissen, M. P., & Reidsma, P. (2021). Participatory assessment of sustainability and resilience of three specialized farming systems. *Ecology and Society, 26*(2), 1-39*.

Pagano, A., Pluchinotta, I., Giordano, R., & Vurro, M. (2017). Drinking water supply in resilient cities: Notes from L'Aquila earthquake case study. *Sustainable Cities and Society, 28*, 435-449*.

Patriarca, R., Bergström, J., Di Gravio, G., & Costantino, F. (2018). Resilience engineering: Current status of the research and future challenges. *Safety Science, 102*, 79-100.

Patriarca, R., Di Gravio, G., & Costantino, F. (2017). A Monte Carlo evolution of the Functional Resonance Analysis Method (FRAM) to assess performance variability in complex systems. *Safety Science, 91*, 49-60*.

Peng, C., Zhen, X., & Huang, Y. (2023). A structured approach for resilience-oriented human performance assessment and prediction in offshore safety-critical operations*. *Ocean Engineering, 287*, 115743.

Peterson, G. (2000). Political ecology and ecological resilience:: An integration of human and ecological dynamics. *Ecological Economics, 35*(3), 323-336*.

Pruyt, E. (2007). *Dealing with uncertainties? Combining system dynamics with multiple criteria decision analysis or with exploratory modelling.* Paper presented at the Proceedings of the 25th International Conference of the System Dynamics Society.

Pumpuni-Lenss, G., Blackburn, T., & Garstenauer, A. (2017). Resilience in complex systems: an agent-based approach. *Systems Engineering, 20*(2), 158-172*.

Rasoulkhani, K., Mostafavi, A., Reyes, M. P., & Batouli, M. (2020). Resilience planning in hazards-humans-infrastructure nexus: A multi-agent simulation for exploratory assessment of coastal water supply infrastructure adaptation to sea-level rise. *Environmental Modelling & Software, 125*, 1-16*.

Ross, A. R., & Chang, H. (2021). Modeling the system dynamics of irrigators' resilience to climate change in a glacier-influenced watershed. *Hydrological Sciences Journal, 66*(12), 1743-1757*.

Saltelli, A., Aleksankina, K., Becker, W., Fennell, P., Ferretti, F., Holst, N., . . . Wu, Q. (2019). Why so many published sensitivity analyses are false: A systematic review of sensitivity analysis practices. *Environmental Modelling & Software, 114*, 29-39.

Salvia, R., & Quaranta, G. (2015). Adaptive cycle as a tool to select resilient patterns of rural development. *Sustainability, 7*(8), 11114-11138*.

San Martín, C., Herrera, H., Paas, W., Reidsma, P., Kopainsky, B., Bertolozzi-Caredio, D., & Soriano, B. (2021). *System dynamics assessment to strengthen resilience and sustainability of farming systems. A participatory approach*. Paper presented at the European Association of Agricultural Economists.

Sendzimir, J., Reij, C. P., & Magnuszewski, P. (2011). Rebuilding resilience in the Sahel: regreening in the Maradi and Zinder regions of Niger. *Ecology and Society, 16*(3), 1-29*.

Shi, Y., Zhai, G., Xu, L., Zhou, S., Lu, Y., Liu, H., & Huang, W. (2021). Assessment methods of urban system resilience: From the perspective of complex adaptive system theory. *Cities, 112*, 1-13*.

Siegenfeld, A. F., Taleb, N. N., & Bar-Yam, Y. (2020). What models can and cannot tell us about COVID-19. *Proceedings of the National Academy of Sciences, 117*(28), 16092-16095.

Simonovic, S. P., & Peck, A. (2013). Dynamic resilience to climate change caused natural disasters in coastal megacities quantification framework. *British Journal of Environment and Climate Change, 3*(3), 378-401*.

Singh, R., Reed, P. M., & Keller, K. (2015). Many-objective robust decision making for managing an ecosystem with a deeply uncertain threshold response. *Ecology and society, 20*(3), 1-33*.
30

* Papers used in the review